# HIDE-AND-SEEK WITH DIRECTIONAL SENSING

ALESSANDRO BORRI[1], SHAUNAK D. BOPARDIKAR[2], JOAO P. HESPANHA[2], MARIA D. DI BENEDETTO[1]

ABSTRACT. We consider a game played between a hider, who hides a static object in one of several possible positions in a bounded planar region, and a searcher, who wishes to reach the object by querying sensors placed in the plane. The searcher is a mobile agent, and whenever it physically visits a sensor, the sensor returns a random direction, corresponding to a half-plane in which the hidden object is located. We first present a novel search heuristic and characterize bounds on the expected distance covered before reaching the object. Next, we model this game as a large-dimensional zero-sum dynamic game and we apply a recently introduced randomized sampling technique that provides a probabilistic level of security to the hider. We observe that, when the randomized sampling approach is only allowed to select a very small number of samples, the cost of the heuristic is comparable to the security level provided by the randomized procedure. However, as we allow the number of samples to increase, the randomized procedure provides a higher probabilistic security level.

## 1. INTRODUCTION

Games of hide-and-seek have been among folklore over the past several centuries. In modern times, these games find relevance in modeling problems of protecting high-valued assets or seeking out potential threats. This paper is concerned with one such game between a hider, who hides a static object among candidate points in a region, and a seeker, who wishes to reach the object by covering minimum distance, while making use of information about the object's location obtained from a few sensors deployed in the region.

A version of this problem without sensor measurements was addressed in our earlier work [BBHPD10], wherein we formalized the problem in the form of a static matrix game. We provided a randomized approach to obtain probabilistic levels of security for the hider. In this paper, we utilize the measurements from additional sensors, which effectively make this problem a dynamic game. In addition, we also analyze a novel heuristic for the seeker, and compare its performance to the randomized method.

**Related Work.** Search theory has received a lot of attention over the past several decades. Classic works [Sto07], [Was02] address several problems involving the search of static as well as moving objects, under various assumptions on sensing abilities of the searcher. [BP02] proposes a search-theoretic approach based on rate of return maps to develop cooperative search plans for uninhabited air vehicles to detect stationary targets.

The problem considered in this paper bears similarities to the area of acoustic source localization, which involves estimating the location of a source using time-of-arrival measurements (see [AKLV04], [BSL08]). Recently, [KSIH10] proposed protocols to route one or more unmanned aerial vehicles to collect time-of-arrival measurements from sensors deployed in an environment. [SMH03] presented an efficient approach to detect an object in a polygonal environment. One straightforward strategy to reach the object without the use of any sensor measurement is to visit every candidate point via the shortest path through all candidate points. However, the computational complexity of determining the shortest path scales exponentially with the number of candidate points [Bel62].

This material is based upon work supported in part by ARO MURI Grant number W911NF0910553, and in part by the Center of Excellence for Research DEWS, University of L'Aquila, Italy.

**Contributions.** We consider a game played between a hider and a searcher. A static object, such as a treasure, is placed by the hider at one of $m$ given points, distributed independently and uniformly in a square region $\mathcal{E} \subset \mathbb{R}^2$. We assume that $s(m)$ directional sensors are deployed in $\mathcal{E}$ so as to minimize the maximum distance from any point in $\mathcal{E}$ [SD96]. Each sensor has an associated line and returns a binary information about the half-plane in which the treasure lies, without measurement error. Such a model arises when an array of microphones is used at each location [VMRL03]. The goal for a searcher, assumed to move with bounded maximum speed and to have a simple integrator dynamics, is to reach the treasure in the shortest possible time (or equivalently, by covering the shortest possible distance), from a given starting point. On the other hand, the hider's goal is to make the searcher travel as much distance as possible until the treasure is reached.

We first present a Divide-and-Search heuristic which involves: (i) moving to the sensor closest to the centroid of the sub-region that contains the treasure, (ii) updating the sub-region that contains the treasure based on the measurement received, and (iii) performing an exhaustive search through all the remaining candidate points, when no more sensor locations are present in the updated sub-region that contains the treasure. We provide a novel upper bound on the expected distance covered by the searcher and characterize the number of sensors required by this heuristic. The upper bound on the expected distance scales as $\log(m)$, which is a significant improvement as compared to following an open-loop approach of following the shortest path through the candidate points, the length of which scales as $\sqrt{m}$.

We then model this game as a zero-sum feedback matrix game and apply the randomized approach introduced in [BBHPD10] for static matrix games, and extended in [BH11] to partial information dynamic games. The hider samples columns from a matrix, which represent different policies of the searcher, and computes a sampled security level and a policy to play the actual game. We first establish monotonicity of the security level of the game as a function of the number of policies of the searcher considered by the hider. A comparison of the security levels obtained through the randomized method with the lengths of the paths produced by the heuristic procedure reveals that if the hider becomes aware that the seeker is using the heuristic procedure she can select (deterministic) locations for the treasure that will lead to very long searches by the seeker. In contrast, the randomized approach produces search strategies that make the task of the hider more challenging, effectively forcing her to use randomized hiding strategies. However, it is important to note that the randomized approach generates policies that are optimized for a specific geometry of the points and can be fragile with respect to changes in the positions of the points away from the positions for which the game was sampled.

**Organization.** This paper is organized as follows. The problem formulation is presented in Section 2. Preliminary results are included in Section 3. The Divide-and-Search heuristic is presented and analyzed in Section 4. The formulation of this problem in the form of a dynamic game is presented in Section 5. Simulations of the heuristic and the randomized approach are included in Section 6.

## 2. Problem formulation

Consider a problem in which a seeker has to find a treasure that is placed at one of $m$ given points, distributed independently and uniformly in a square region $\mathcal{E} \subset \mathbb{R}^2$ with area denoted by Area($\mathcal{E}$). We consider deploying $s(m)$ directional sensors in $\mathcal{E}$ so that the maximum distance of any point in $\mathcal{E}$ from any sensor is minimized. In the following, each point will be regarded in terms of the corresponding index $i \in \mathcal{P} := \{1, 2, ..., m\}$ for each candidate treasure point, $i \in \mathcal{S} := \{m+1, ..., m+s\}$ for each sensor point and $i = 0$ for the searcher initial position, that is assumed to be set at the centre of the square region.

Each sensor $i$ has an associated straight line defined by $n_i'(p - p_i) = 0$, $i \in \mathcal{S}$, where the unit vector $n_i$ (normal to the straight line) is given and known, and $p_i$ belongs to the line; we assume that $n_i'(p_j - p_i) \neq 0$ for all $i \in \mathcal{S}$, $j \in \mathcal{P}$, $i \neq j$, and that the orientations of the unit vectors $n_i$ are distributed independently and uniformly in $[0, 2\pi]$. Whenever a sensor is visited, it returns a binary information about the half-plane in which the treasure lies, as shown in Figure 1. More formally, a seeker visiting a sensor point $i$ will receive the observation $y = \text{sgn}(n_i'(p_\Theta - p_i)) \in \{-1, 1\}$, where $p_\Theta$ denotes the treasure position. This will allow us to restrict the time horizon to at most $m + s$ points $\mathcal{T} = \{0, 1, 2, ..., m + s\}$, after which the treasure is certainly found.

We shortly introduce a bit of extra notation to better describe the strategy of the seeker. At time $t \in \mathcal{T}$, let $\mathbf{x}(t)$ be current location of the pursuer, with $\mathbf{x}(0) = 0$ and $\mathbf{x}(t) \in \mathcal{P} \cup \mathcal{S}$ for $t \geq 1$. The position $\mathbf{x}(t + 1)$ is decided at time $t$ by a motion control action $\mathbf{u}(t)$, namely $\mathbf{x}(t + 1) = \mathbf{u}(t)$; the pursuer visits the node $p_{\mathbf{x}(t)}$ and gets the measurement $\mathbf{y}(t) = \text{sgn}(n'_{\mathbf{x}(t)}(p_\Theta - p_{\mathbf{x}(t)})) \in \{-1, 1\}$, where $\Theta \in \mathcal{P}$ is the treasure position.

Let $X_k = \{\mathbf{x}(t)\}_{t=0}^{k}$ and $Y_k = \{\mathbf{y}(t)\}_{t=0}^{k}$ be the sequences of visited nodes and collected observations by the seeker, respectively, up to some time $k \geq 0$, with $\mathbf{y}(0) := \emptyset$. For the sake of simplicity, we assume we can write $X_{k+1} = X_k \cup \{\mathbf{x}(k + 1)\}$ and $Y_{k+1} = Y_k \cup \{\mathbf{y}(k + 1)\}$ with a slight abuse of notation.

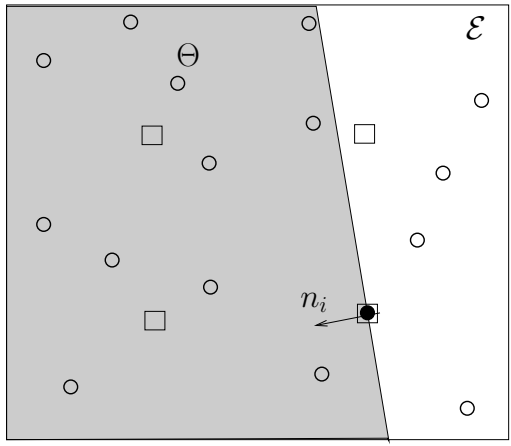

FIGURE 1. Problem formulation. A circle denotes a candidate treasure point, a square denotes a sensor and the dot denotes the searcher. In this figure, the searcher is at a sensor and has received a measurement corresponding to the treasure $\Theta$ being at one of the circles in the shaded region.

## 3. Preliminaries

In this section, we summarize some preliminary results which we will use in the rest of the paper.

### 3.1. Geometric Preliminaries.
The first result by [SD96] gives an upper bound on the minimum of the $s$ distances from any point $\mathbf{z} \in \mathcal{E}$.

**Lemma 3.1.** *The rectangular heuristic from* [SD96] *to place the $s$ sensors satisfies*

$$\min_{i \in \{1, ..., s\}} \|\mathbf{z} - p_i\| \leq \frac{1}{\sqrt{2s}} \sqrt{\text{Area}(\mathcal{E})}$$

*for any $\mathbf{z} \in \mathcal{E}$.*

The next result is well-known in computational geometry.

**Proposition 3.2** (Minimum enclosing rectangle). *Given a convex polygon with area $A$, the minimum area-enclosing rectangle has area upper bounded by $2A$.*

The next result provides an upper bound on the length of the shortest path through all points in a convex polygon which is contained inside $\mathcal{E}$.

**Lemma 3.3** (Shortest path length bound). *The length of the shortest path through $n$ points that are inside a convex polygon of area $A$ which is contained inside $\mathcal{E}$ is upper bounded by*

$$2\sqrt{An} + \frac{9}{\sqrt{2}}\sqrt{\mathrm{Area}(\mathcal{E})}.$$

*Proof of Lemma 3.3.* We provide a constructive proof as follows. First, consider the minimum area-enclosing rectangle of the given convex polygon. By Proposition 3.2, this rectangle has area upper bounded by $2A$. Now, let the smaller side of this rectangle have length equal to $B$ and the larger side have length equal to $H$. Define, $h := H/B = 2A/B^2$. Now using Lemma II.3 from [BSBH10], there exists a path that starts from the smaller side, passes through each of the $n$ points exactly once and terminates on the opposite side of the rectangle, and has length upper bounded by

$$B(\sqrt{2hn} + h + 5/2) = 2\sqrt{An} + H + \frac{5}{2}B \leq 2\sqrt{An} + \frac{7}{\sqrt{2}}\sqrt{\mathrm{Area}(\mathcal{E})},$$

where we used the fact that both $B$ and $H$ can be at most equal to $\sqrt{2\,\mathrm{Area}(\mathcal{E})}$. The result now follows since the vehicle would cover a distance of at most $\sqrt{2\,\mathrm{Area}(\mathcal{E})}$ to reach any point on the smaller side of the minimum area-enclosing rectangle of the given convex polygon. $\qquad\square$

### 3.2. Randomized sampling for large zero-sum games.
In [BBHPD10], we introduced the *Sampled Saddle-Point* (SSP) algorithm, in which players sample sub-matrices from the original $M \times N$ matrix $A$, solve smaller games and utilize the saddle-point policies so obtained against each other. The SSP algorithm can be summarized as follows.

Let $\mathcal{B}^{k \times l}$ denote the set of $k \times l$ left-stochastic $(0, 1)$-matrices (i.e., matrices whose entries belong to the set $\{0, 1\}$ and whose columns add up to one).

---

**Algorithm 1: Sampled Saddle-Point**

**For $P_1$:**
- Select random matrices $\Gamma_1 \in \mathcal{B}^{M \times m_1}$, $\Pi_1 \in \mathcal{B}^{N \times n_1}$
- Compute sub-matrix: $A_1 = \Gamma_1' A \Pi_1$
- Security policy: $y_1^* \in \mathrm{argmin}_{y_1 \in \mathcal{S}_{m_1}} \max_{z \in \mathcal{S}_{n_1}} y_1' A_1 z$
- Security value: $\bar{V}(A_1) = \max_{z \in \mathcal{S}_{n_1}} y_1^{*\prime} A_1 z$

**For $P_2$:**
- Select random matrices $\Gamma_2 \in \mathcal{B}^{M \times m_2}$ and $\Pi_2 \in \mathcal{B}^{N \times n_2}$
- Compute sub-matrix: $A_2 = \Gamma_2' A \Pi_2$
- Security policy: $z_2^* \in \mathrm{argmax}_{z_2 \in \mathcal{S}_{n_2}} \min_{y \in \mathcal{S}_{m_1}} y' A_2 z_2$
- Security value: $\underline{V}(A_2) = \min_{y \in \mathcal{S}_{m_1}} y' A_2 z_2^*$

**Play sampled policies:** $y^* := \Gamma_1 y_1^*$, $z^* := \Pi_2 z_2^*$

**Output:** $y^{*\prime} A z^*$

---

The SSP algorithm is *$\epsilon$-secure for player $P_1$ with confidence $1 - \delta$* if

$$(3.1) \qquad \mathbb{P}_{\Gamma_1, \Pi_1, \Gamma_2, \Pi_2}\left(y^{*\prime} A z^* \leq \bar{V}(A_1) + \epsilon\right) \geq 1 - \delta.$$

The subscript in the probability measure $\mathbb{P}$ emphasizes which random variables define the events that are being measured. To provide guarantees for specific policies/values, the following notions of security that refer to specific policies/values are introduced. The policy $y^*$ with value $\bar{V}(A_1)$ is *$\epsilon$-secure for player $P_1$ with confidence $1 - \delta$* if

$$(3.2) \qquad \mathbb{P}_{\Gamma_1, \Pi_1}\left(y^{*\prime} A z^* \leq \bar{V}(A_1) + \epsilon \,\middle|\, y^*, \,\bar{V}(A_1)\right) \geq 1 - \delta.$$

The following result provides a bound on the size of the sub-matrices for the players that guarantees $\epsilon$-security with $\epsilon = 0$.

**Theorem 3.4** (Game independent bounds). *Suppose that the matrices $\Gamma_1, \Pi_1, \Gamma_2, \Pi_2$ are statistically independent. Then, If $\Pi_1$ and $\Pi_2$ have identically distributed columns and*

$$n_1 = \left\lceil \frac{m_1 + 1}{\delta} - 1 \right\rceil \bar{n}_2,$$

*for some $\bar{n}_2 \geq n_2$, then the SSP algorithm is $\epsilon = 0$-secure for $P_1$ with confidence $1 - \delta$.*

*If one further increases $n_1$ to satisfy*

$$n_1 = \left\lceil \frac{1}{\delta} \left( \ln \frac{1}{\beta} + m_1 + 2\sqrt{m_1 \ln \frac{1}{\beta}} \right) \right\rceil \bar{n}_2,$$

*then, with probability[1] higher than $1 - \beta$, the policy $y^*$ with value $\bar{V}(A_1)$ is $\epsilon = 0$-secure for $P_1$ with confidence $1 - \delta$.*

Suppose that, due to computational limitations, player $P_1$ cannot satisfy the bounds in Theorem 3.4 to obtain $\epsilon = 0$-security for a given level of confidence $1 - \delta$. One option to overcome this difficulty and maintain the same high level of confidence would be to accept a larger value for $\epsilon$.

The following is an a-posteriori procedure for $P_1$. Let $e_j$ denote the $j$th element of the canonical basis of $\mathbb{R}^{k_1}$.

---
**Algorithm 2: A-posteriori procedure**

- Pick values for $m_1$ and $n_1$
- Determine $y^*$ and $\bar{V}(A_1)$ using SSP algorithm
- Select random matrix $\bar{\Pi}_1 \in \mathcal{B}^{N \times k_1}$ using distribution of $\Pi_1$

    **Output:** $\bar{v} = \max_{j \in \{1, \dots, k_1\}} y_1^{*\prime} A \bar{\Pi}_1 e_j$

---

The following result provides an a-posteriori guarantee on this procedure.

**Theorem 3.5** (A-posteriori bounds). *Suppose that the matrices $\Gamma_1, \Pi_1, \Gamma_2, \Pi_2$ are statistically independent. If $\Pi_1$ and $\Pi_2$ have identically distributed columns and*

$$(3.3) \qquad k_1 = \left\lceil \frac{1}{\delta} - 1 \right\rceil \bar{n}_2,$$

*for some $\bar{n}_2 \geq n_2$, then the SSP algorithm is $\epsilon$-secure for $P_1$ with confidence $1 - \delta$ for any $\epsilon \geq \bar{v} - \bar{V}(A_1)$.*

*If one further increases $k_1$ to satisfy*

$$k_1 = \left\lceil \frac{\ln(1/\beta)}{\ln(1/(1-\delta))} \right\rceil \bar{n}_2,$$

*then, with probability higher than $1 - \beta$, the policy $y^*$ with value $\bar{V}(A_1)$ is $\epsilon$-secure for $P_1$ with confidence $1 - \delta$.*

## 4. THE DIVIDE-AND-SEARCH HEURISTIC

We now present and analyze a novel heuristic for the searcher. This heuristic involves computing the centroid of a convex polygonal region at every iteration, and then moving to the sensor closest to the centroid. At every iteration, the polygonal region is updated using the measurement from the sensor. If no such sensor exists, then the searcher performs an exhaustive search over the remaining candidate points in the region.

Formally, the *centroid* Cent($\mathcal{Q}$) of a convex region $\mathcal{Q} \subset \mathbb{R}^2$ is defined as the unique point $q$ in $\mathcal{Q}$ that minimizes $\int_{\mathcal{Q}} \|\mathbf{z} - q\|^2 \, d\mathbf{z}$. As per our measurement model, given an unknown treasure position $\Theta$, any sensor $x(t) \in \mathcal{S}$

---

[1] The confidence level $\beta$ for $P_1$ refers solely to the extraction of the matrix $\Pi_1$ and holds for any given matrix $\Gamma_1$.

visited at time $t$, and a measurement $y(t) = \text{sgn}(n'_{x(t)}\left(p_\Theta - p_{x(t)}\right))$, let $H(x(t), y(t))$ denote the half-plane which contains the treasure.

Let $K$ denote the number of measurements taken by the searcher. Then, the heuristic described in Algorithm 3.

---

**Algorithm 3: Divide-and-Search**

---
**Assumes:** $\mathcal{E}_0 := \mathcal{E}$.
**For** $t = 1, \ldots, K$,

- Go to the sensor $x(t)$ closest to $\text{Cent}(\mathcal{E}_{t-1})$
- Obtain measurement $y(t)$
- Determine $\mathcal{E}_t = \mathcal{E}_{t-1} \cap H(x(t), y(t))$

**end for**
Move on the shortest path through all targets in $\mathcal{E}_K$.

---

Figure 2 shows snapshots of a numerical implementation of the heuristic, with a high number of nodes. The following is a useful property of Algorithm 3.

**Lemma 4.1** (Upper bound on $\text{Area}(\mathcal{E}_K)$). *Assume that the sensors are placed using the rectangular heuristic by [SD96]. Then, at the end of Algorithm 3,*

$$(4.1) \qquad \mathbb{E}[\text{Area}(\mathcal{E}_K)] \leq \left(\left(\frac{2}{3}\right)^K + \frac{1}{\sqrt{2s}}\right)\text{Area}(\mathcal{E}),$$

*where the expectation is over the distribution of the measurements obtained by the sensors.*

*Proof of Lemma 4.1.* From Algorithm 3, it is immediate that each $\mathcal{E}_t$ is convex. Consider the region $\mathcal{E}_{t-1}$ as shown in Figure 3. We obtain a measurement $y(t)$ at a sensor closest to $\text{Cent}(\mathcal{E}_{t-1})$, as shown in Figure 3. Let $l$ denote a line that passes through $\text{Cent}(\mathcal{E}_t)$ and which is perpendicular to $n_{x(t)}$. Using the centroid property [BH02] that any line through the centroid of a convex region divides the region into two parts such that the area of each part is at most two-thirds of the area of the original convex region, we obtain

$$\text{Area}(\mathcal{E}_t) \leq \frac{2}{3}\text{Area}(\mathcal{E}_{t-1}) + d_t\sqrt{\text{Area}(\mathcal{E})},$$

where $d_t$ is the distance of the closest sensor to $\text{Cent}(\mathcal{E}_{t-1})$ and the second term is the area of the rectangular strip between the two lines. Using Lemma 3.1,

$$\text{Area}(\mathcal{E}_t) \leq \frac{2}{3}\text{Area}(\mathcal{E}_{t-1}) + \frac{1}{\sqrt{2s}}\text{Area}(\mathcal{E})$$

$$\Rightarrow \mathbb{E}[\text{Area}(\mathcal{E}_t)] \leq \frac{2}{3}\mathbb{E}[\text{Area}(\mathcal{E}_{t-1})] + \frac{1}{\sqrt{2s}}\text{Area}(\mathcal{E}),$$

where the expectation is with respect to the sensor measurements. By using this inequality recursively, and from the fact that $1 - (2/3)^K < 1$, we obtain (4.1). $\qquad \square$

We now present the main result of this section, which provides an upper bound on the expected distance covered by the searcher using the Divide-and-Search heuristic, and also characterizes the number of sensors required. Let $D$ denote the distance covered by the searcher until the treasure is found.

**Theorem 4.2** (Performance of Algorithm 3). *Suppose that:*

(1) *the number of candidate treasure locations $m$ and the number of sensors $s$ satisfy $s = \lceil m/(\ln(m))^2 \rceil$;*
(2) *the sensors are located using the rectangular heuristic from [SD96].*

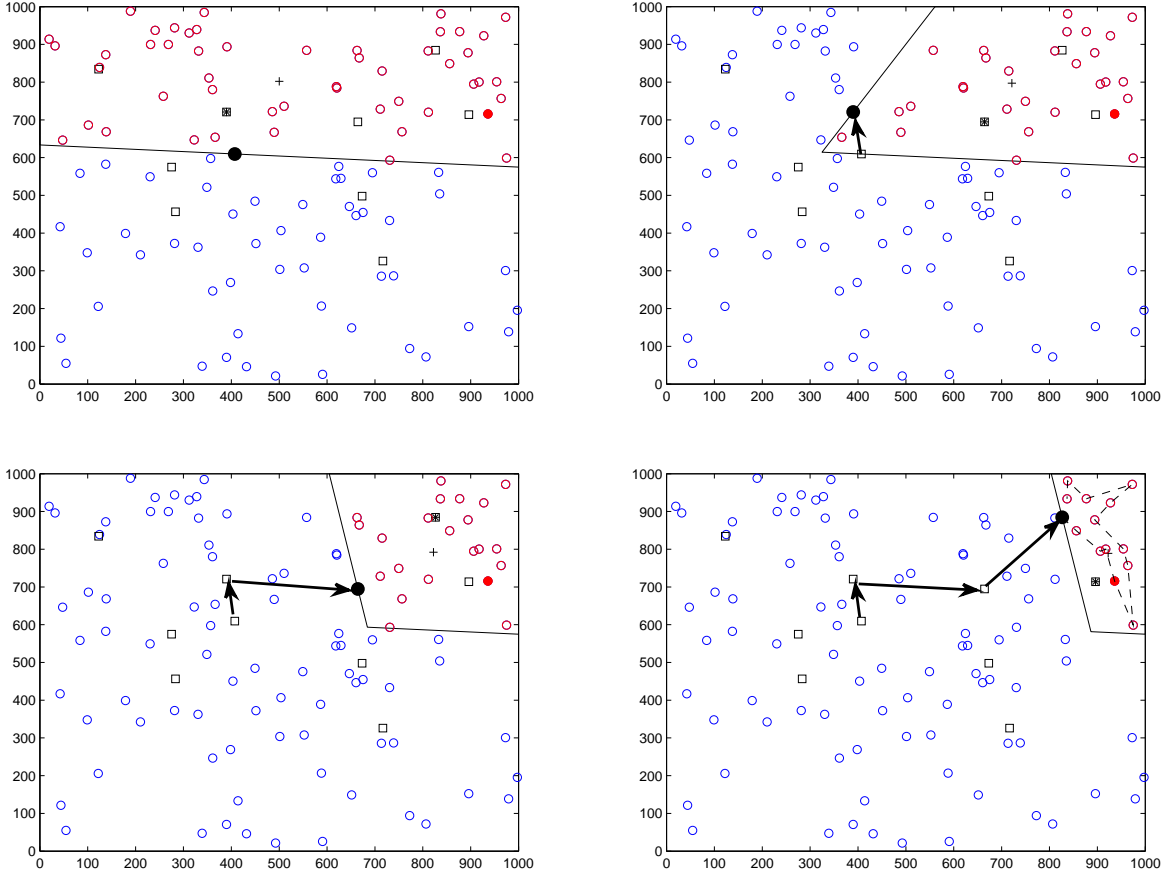

FIGURE 2. The Divide-and-Search heuristic. The big solid dot is the searcher, the blue (light) circles denote target points wherein the treasure is not present, the red (dark) circles denote the candidate treasure locations, the solid red dot denotes the exact location of the treasure, the + denotes the centroid of the region $\mathcal{E}_t$ at each iteration, the squares denote the sensors. The shortest path is illustrated by a dashed line in the final figure.

Then, the expected value of $D$ using Algorithm 3 satisfies

$$\frac{1}{2}\sqrt{\mathrm{Area}(\mathcal{E})} \leq \mathbb{E}[D] \leq \frac{\sqrt{2}\ln(\sqrt{2}\ln(3/2)\sqrt{m}) + \sqrt{2} + \sqrt{2}\ln(3/2)\ln(m)}{\ln(3/2)}\sqrt{\mathrm{Area}(\mathcal{E})},$$

where the expectation is with respect to the joint distribution of the measurements and the candidate treasure locations.

*Proof of Theorem 4.2.* Let $\mathrm{dist}(K)$ denote the distance covered by the vehicle by using Algorithm 3. Clearly, $D \leq \mathrm{dist}(K)$, which is a random variable which is upper bounded by the sum of two terms. The first is $K$ times the diameter of $\mathcal{E}$, and the second is the length of the shortest path through the remaining $m_{\mathrm{rem}}$ candidate locations in $\mathcal{E}_K$. Using Lemma 3.3,

$$\mathrm{dist}(K) \leq K\sqrt{2\,\mathrm{Area}(\mathcal{E})} + 2\sqrt{m_{\mathrm{rem}}\,\mathrm{Area}(\mathcal{E}_K)} + \frac{9}{\sqrt{2}}\sqrt{\mathrm{Area}(\mathcal{E})}.$$

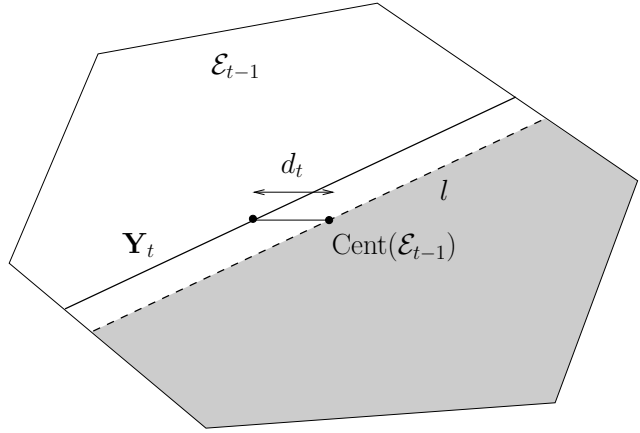

FIGURE 3. Illustrating proof of Lemma 4.1. The dashed line is perpendicular to $n_{x(t)}$ and passes through the centroid of $\mathcal{E}_{t-1}$. By the centroid property, the area of the shaded region is at most two-thirds of the area of $\mathcal{E}_{t-1}$.

Taking expectation with respect to the joint distribution of the measurements from the sensors and the target points,

$$\mathbb{E}[D] \le (\frac{2K+9}{\sqrt{2}})\sqrt{\text{Area}(\mathcal{E})} + 2\mathbb{E}[\sqrt{m_{\text{rem}}\,\text{Area}(\mathcal{E}_K)}] =$$

$$= (\frac{2K+9}{\sqrt{2}})\sqrt{\text{Area}(\mathcal{E})} + 2\mathbb{E}[\mathbb{E}[\sqrt{m_{\text{rem}}\,\text{Area}(\mathcal{E}_K)}\,|\,\text{Area}(\mathcal{E}_K)]] \le$$

$$\le (\frac{2K+9}{\sqrt{2}})\sqrt{\text{Area}(\mathcal{E})} + 2\mathbb{E}[\sqrt{\text{Area}(\mathcal{E}_K)}\sqrt{\mathbb{E}[m_{\text{rem}}\,|\,\text{Area}(\mathcal{E}_K)]}]$$

where the second step follows from the law of iterated expectations, and the third step is due to the application of Jensen's inequality [Bre92] to the square-root function. In the above inequality, the outer expectation is with respect to the measurements while the inner expectation is with respect to the target points. Now, conditioned on the value of $\text{Area}(\mathcal{E}_K)$, the only information about the candidate locations that can be obtained is the existence of the treasure inside $\mathcal{E}_K$. The remaining $m-1$ locations are distributed independently and thus, the expected number of target points inside $\mathcal{E}_K$ satisfies

$$\mathbb{E}[m_{\text{rem}}\,|\,\text{Area}(\mathcal{E}_K)] = 1 + \frac{m-1}{\text{Area}(\mathcal{E})}\,\text{Area}(\mathcal{E}_K) \le 1 + \frac{m\,\text{Area}(\mathcal{E}_K)}{\text{Area}(\mathcal{E})}.$$

On substitution, and since $\sqrt{\text{Area}(\mathcal{E}_K)} \le \sqrt{\text{Area}(\mathcal{E})}$, we have

$$\mathbb{E}[D] \le (\frac{2K+9}{\sqrt{2}} + 2)\sqrt{\text{Area}(\mathcal{E})} + \frac{2\sqrt{m}}{\sqrt{\text{Area}(\mathcal{E})}}\mathbb{E}[\text{Area}(\mathcal{E}_K)].$$

Using Lemma 4.1, we have

$$(4.2) \qquad \mathbb{E}[D] \le \left(\frac{2K+9}{\sqrt{2}} + 2 + 2\sqrt{m}\left(\frac{2}{3}\right)^K + \sqrt{2}\ln(m)\right)\sqrt{\text{Area}(\mathcal{E})},$$

since by assumption, $s = \lceil m/(\ln(m))^2 \rceil$. Now, the right hand side of the above inequality is minimized with

$$K^* = \ln\left(\sqrt{2}\ln(3/2)\sqrt{m}\right)/\ln(3/2).$$

Substituting this value of $K$ in the right hand side of (4.2), we obtain the right hand side of the first inequality. The left hand side is trivially true because the treasure can be placed at the diametrically opposite end of the region $\mathcal{E}$. $\qquad \square$

*Remark* 4.3 (Number of sensors). Theorem 4.2 implies that with at most $\lceil m/(\ln(m))^2 \rceil$ number of sensors, and with a number of steps $K^*$ which grow logarithmically as a function of $m$, Algorithm 3 may lead to an average cost which is at most $\ln(m)$ times the optimal.

## 5. Formulation as a Matrix Game

Consider an output feedback zero-sum dynamic game where $P_1$ hides a non-moving object (treasure) in one of $m$ points $\{p_i\}_{i=1}^m$ on the plane and $P_2$ has to find the treasure with minimum cost, by traveling from point to point until she finds it, starting at $p_0$. At each step, $P_2$ can either visit one of the $m$ points, trying to guess the treasure position, or visit one of $s(m)$ sensor points $\{p_i\}_{i=m+1}^{m+s}$, to get better information. The game is played over the set of mixed policies:

- $P_1$ chooses a probability distribution $y \in \mathcal{S}_m$ for the treasure over the $m$ points;
- $P_2$ chooses a probability distribution $z \in \mathcal{S}_n$ over the set of all the possible $n$ functions of the available information, mapping the information available to $P_2$ at some time step to the action to be taken at that step. Specifically, the information available consists of the sequence of points visited so far and the measurements collected from the sensors visited. The next action to take consists of the index of the next point to visit.

The game can be formulated as a matrix game with a very large number of columns, where the generic entry $A_{ij}$ corresponds to the Euclidean cost to find the treasure by playing policy $j$ when the treasure is placed in $i$, namely:

$$(5.1) \qquad A_{ij} = -\sum_{k=0}^{k_{ij}^*} \|p_{r_j(X_k, Y_k)} - p_{r_j(X_{k-1}, Y_{k-1})}\|,$$

where $r_j(X_{-1}, Y_{-1}) := 0$ for all $j$ (starting point) and the summation ends at the index $k_{ij}^*$ for which $r_j(X_{k_{ij}^*}, Y_{k_{ij}^*})$ corresponds to the point $i$ where the treasure is hidden. The minus sign in Eq. (5.1) is needed to maintain consistency with the formulation in Subsection 3.2, where $P_1$ is the minimizer. Indeed, $P_1$ hides the treasure to maximize the distance and therefore to minimize the entries of $A$.

The exact computation of the optimal mixed strategies is intractable because the size of the matrix is very large, in general. However, the results regarding the SSP algorithm have a computational complexity that is completely *independent of the size of the game,* which means that we can provide probabilistic guarantees for games with an arbitrarily large number of points.

In this particular game, only the player $P_2$ has a very large number of options, so we can assume that both players consider all possible $m$ locations where $P_1$ can hide the treasure (all rows of $A$), but randomly select only a small number of pursuit strategies to construct their submatrices. This means that $P_2$ will never be surprised since she always considers all options for the actions of $P_1$. However, the player $P_1$ that hides the treasure should respect the bounds provided by Theorems 3.4 and 3.5 to avoid unpleasant surprises.

The choice $m_k = m$ in the SSP algorithm is particularly interesting to apply in games where the matrix is "fat", with many more columns that rows. In these cases, further results can be obtained for the SSP procedure, as shown in Subsection 5.1.

### 5.1. **SSP for "fat" matrix games.**

5.1.1. *Sampled security level $\bar{V}(A_1)$.* In the SSP algorithm, the sampled security level for $P_1$ is a random variable, given by

$$\bar{V}(A_1) = \min_{y \in \mathcal{S}_{m_1}} \max_{z \in \mathcal{S}_{n_1}} y'A_1 z.$$

In the case $m_1 = m$, $\bar{V}(A_1)$ only depends on $n_1$ and on the particular realization of the $n_1$ columns, and thus is a function of the random variable $\Pi_1$. Thus,

$$\bar{V}(A_1) = \min_{y \in S_m} \max_{z \in S_{n_1}} y'A_1 z = \max_{z \in S_{n_1}} y^{*\prime} A_1 z = \max_{j \in \{1,\dots,n_1\}} y^{*\prime} A \Pi_1 e_j.$$

The following preliminary result essentially states that as the number of sampled columns increases, the sampled value $\bar{V}(A_1)$ is expected to increase, or in other words, the probability that $\bar{V}(A_1)$ is less than a fixed quantity decreases monotonically.

**Lemma 5.1.** *For any fixed $x \in \mathbb{R}$, $\mathbb{P}_{\Pi_1}\left(\bar{V}(A_1) \leq x\right)$ is non–increasing with $n_1$. Further, if every column of $A$ is being sampled with a positive probability by the SSP algorithm, then, in the limit as $n_1$ goes to infinity, $\bar{V}(A_1)$ converges almost surely to the original value of the game $V(A)$.*

*Proof of Lemma 5.1.* We begin with the first claim.

$$\mathbb{P}_{\Pi_1}\left(\bar{V}(A_1) \leq x\right) = \mathbb{P}_{\Pi_1}\left(\max_{j \in \{1,\dots,n_1\}} y^{*\prime} A \Pi_1 e_j \leq x\right) =$$

$$= \sum_{\bar{\Pi}_1 \in \mathcal{B}^{N \times n_1}} \mathbb{P}_{\Pi_1}\left(\max_{j \in \{1,\dots,n_1\}} y^{*\prime} A \Pi_1 e_j \leq x | \Pi_1 = \bar{\Pi}_1\right) \cdot \mathbb{P}_{\Pi_1}\left(\Pi_1 = \bar{\Pi}_1\right) =$$

$$= \sum_{\bar{\Pi}_1 \in \mathcal{B}^{N \times n_1}} \mathbb{P}_{\Pi_1}\left(\max_{j \in \{1,\dots,n_1\}} y^{*\prime} A \Pi_1 e_j \leq x | \Pi_1 = \bar{\Pi}_1\right) \cdot \mathbb{P}_{\Pi_1}\left(\Pi_1 = \bar{\Pi}_1\right),$$

where $e_j$ is the $j$th element of the canonical basis of $\mathbb{R}^{n_1}$.

Now, consider the concatenated matrix $\tilde{\Pi}_1 := [\Pi_1 \, e]$, where $e \in \mathcal{B}^{N \times 1}$. Then,

$$\mathbb{P}_{\tilde{\Pi}_1}\left(\bar{V}(A_1) \leq x\right) = \sum_{\Pi} \mathbb{P}_{\tilde{\Pi}_1}\left(\max_{j \in \{1,\dots,n_1+1\}} y^{*\prime} A \tilde{\Pi}_1 e_j \leq x | \tilde{\Pi}_1 = \Pi\right) \cdot \mathbb{P}_{\tilde{\Pi}_1}\left(\tilde{\Pi}_1 = \Pi\right) =$$

$$= \sum_{\Pi} \mathbb{P}_{\Pi_1,e}\left(\max_{j \in \{1,\dots,n_1+1\}} y^{*\prime} A [\Pi_1 \, e] e_j \leq x | \Pi_1 = \bar{\Pi}_1, e = \bar{e}\right) \cdot \mathbb{P}_{\Pi_1}\left(\Pi_1 = \bar{\Pi}_1\right) \cdot \mathbb{P}_e\left(e = \bar{e}\right) \leq$$

$$\leq \sum_{\bar{\Pi}_1} \mathbb{P}_{\Pi_1,e}\left(\max_{j \in \{1,\dots,n_1\}} y^{*\prime} A \Pi_1 e_j \leq x | \Pi_1 = \bar{\Pi}_1, e = \bar{e}\right) \cdot \mathbb{P}_{\Pi_1}\left(\Pi_1 = \bar{\Pi}_1\right) \cdot \sum_{\bar{e}} \mathbb{P}_e\left(e = \bar{e}\right) =$$

$$= \sum_{\bar{\Pi}_1} \mathbb{P}_{\Pi_1}\left(\max_{j \in \{1,\dots,n_1\}} y^{*\prime} A \Pi_1 e_j \leq x | \Pi_1 = \bar{\Pi}_1\right) \cdot \mathbb{P}_{\Pi_1}\left(\Pi_1 = \bar{\Pi}_1\right) =$$

$$= \mathbb{P}_{\Pi_1}\left(\bar{V}(A_1) \leq x\right).$$

Thus, the first claim stands proved.

For the second claim, let $c_j$ denote the $j$th element of the canonical basis of $\mathbb{R}^N$, and let $C_1$ denote the first column of $\Pi_1$. Now, the probability that $c_j$ is sampled is given by $\mathbb{P}_{\Pi_1}(c_j \in \text{Range}(\Pi_1))$. By assumption, $\mathbb{P}_{C_1}(C_1 = c_j) > 0$, $\forall j \in \{1,\dots,n\}$. This implies that $\mathbb{P}_{C_1}(C_1 \neq c_j) < 1$, $\forall j \in \{1,\dots,N\}$. For any $j$, the probability of $c_j$ being drawn in $n_1$ independent trials is

$$\mathbb{P}_{\Pi_1}(c_j \in \text{Range}(\Pi_1)) = 1 - \mathbb{P}_{\Pi_1}(c_j \notin \text{Range}(\Pi_1)) =$$

$$= 1 - \mathbb{P}_{C_1}\left(C_1 \neq c_j\right)^{n_1}.$$

Since $\mathbb{P}_{C_1}\left(C_1 \neq c_j\right) \in (0,1)$ by assumption, the probability that $c_j$ does not get sampled goes to zero exponentially with $n_1$. This implies convergence in probability of $\bar{V}(A_1)$ to $V(A)$, and almost sure convergence is implied by the use of Borel-Cantelli lemma [Res98]. □

We now recall that the quantile $X_\alpha$ of a random variable $X$ is the inverse function of the cumulative distribution function and returns the value below which random draws from the given distribution would fall ($\alpha \times 100$)-percent of the time. For any discrete random variable $X$ and a real number $\alpha \in [0, 1]$, the quantile is defined as

$$X_\alpha = \inf \{x \in \mathbb{R} : \mathbb{P}(X \leq x) \geq \alpha\},$$

and is a non-decreasing function of $\alpha$.

The next result shows that, for any fixed $\alpha$, the quantile of the sampled value $\bar{V}(A_1)$ is a non-decreasing function of $n_1$.

**Proposition 5.2.** *For any $\alpha \in [0, 1]$, the quantile $\bar{V}_\alpha(A_1(n_1))$ of the sampled value of the game $\bar{V}(A_1(n_1))$ is a non-decreasing function of $n_1$. Moreover, in the limit as $n_1$ goes to infinity, $\bar{V}_\alpha(A_1(n_1))$ converges almost surely to the original value of the game, for every $\alpha \in [0, 1]$.*

*Proof of Proposition 5.2.* For any $\alpha$ and $n_1$, consider the quantiles

$$\bar{V}_\alpha(A_1(n_1)) = \inf \{x \in \mathbb{R} : \mathbb{P}(\bar{V}(A_1(n_1)) \leq x) \geq \alpha\},$$
$$\bar{V}_\alpha(A_1(n_1 + 1)) = \inf \{x \in \mathbb{R} : \mathbb{P}(\bar{V}(A_1(n_1 + 1)) \leq x) \geq \alpha\}.$$

By definition of the quantile, we have

$$\mathbb{P}(\bar{V}(A_1(n_1)) \leq \bar{V}_\alpha(A_1(n_1))) \geq \alpha,$$
$$\mathbb{P}(\bar{V}(A_1(n_1)) \leq \bar{V}_\alpha(A_1(n_1)) - \epsilon) < \alpha, \text{ for all } \epsilon > 0.$$

Now, assume that

$$\bar{V}_\alpha(A_1(n_1 + 1)) = \bar{V}_\alpha(A_1(n_1)) - \delta < \bar{V}_\alpha(A_1(n_1)),$$

for some $\delta > 0$, i.e., the quantile of $\bar{V}$ is strictly decreasing. Then,

$$\mathbb{P}(\bar{V}(A_1(n_1 + 1)) \leq \bar{V}_\alpha(A_1(n_1 + 1))) = \mathbb{P}(\bar{V}(A_1(n_1 + 1)) \leq \bar{V}_\alpha(A_1(n_1)) - \delta) \leq$$
$$\leq \mathbb{P}(\bar{V}(A_1(n_1)) \leq \bar{V}_\alpha(A_1(n_1)) - \delta) < \alpha,$$

where we used Lemma 5.1. But, by the definition of quantile,

$$\mathbb{P}(\bar{V}(A_1(n_1 + 1)) \leq \bar{V}_\alpha(A_1(n_1 + 1))) \geq \alpha,$$

which is a contradiction, and therefore $\bar{V}_\alpha(A_1(n_1))$ is a non-decreasing function of $n_1$.

Furthermore, for all $\alpha > 0$, we have

$$\lim_{n_1 \to \infty} \bar{V}_\alpha(n_1) = \lim_{n_1 \to \infty} \inf \{x \in \mathbb{R} : \mathbb{P}(\bar{V}(A_1(n_1)) \leq x) \geq \alpha\} =$$
$$= \inf \{x \in \mathbb{R} : \mathbb{P}(V(A) \leq x) \geq \alpha\} = V(A)$$

where the limit and the infimum commute due to convergence of $\bar{V}(A_1(n_1))$ established in Lemma 5.1. □

5.1.2. *Outcome of the game $\bar{v}$.* We now consider the expression of the a-posteriori outcome $\bar{v}$ (see Subsection 3.2), which is also a random variable:

$$\bar{v}(n_1, k_1, \Pi_1, \bar{\Pi}_1) = \max_{j \in \{1, \dots, k_1\}} y^{*\prime} A \bar{\Pi}_1 e_j,$$

where $k_1$ is given by Eq. (3.3), and we consider its quantile

$$\bar{v}_\alpha(n_1, k_1) := \inf \{x \in \mathbb{R} : \mathbb{P}_{\Pi_1, \bar{\Pi}_1}(\bar{v}(n_1, k_1, \Pi_1, \bar{\Pi}_1) \leq x) \geq \alpha\}.$$

Analogous to the quantile of the sampled value of the game $\bar{V}(A_1)$, it can be shown that the quantile of $\bar{v}$ is monotonically non-decreasing as in the following result.

**Proposition 5.3.** *For any $\alpha \in [0,1]$, the quantile $\bar{v}_\alpha$ of the a-posteriori outcome of the game $\bar{v}$ is a non-decreasing function of $k_1$, for any fixed $n_1$. Moreover, in the limit as $n_1$ and $k_1$ tend to infinity, $\bar{v}_\alpha$ converges in distribution to the original value of the game for any $\alpha \in [0,1]$, i.e.*

$$\lim_{n_1, k_1 \to \infty} \bar{v}_\alpha(n_1, k_1, \Pi_1, \bar{\Pi}_1) = V(A).$$

*Remark* 5.4. Note that, since for both the bounds provided by Theorems 3.4 and 3.5 we have $\lim_{\delta \to 0} k_1 = +\infty$, the previous results can be restated in terms of $\delta$; in particular $\bar{v}_\alpha(n_1, k_1)$ is non-increasing with $\delta$ for any fixed $n_1$, and in the limit ($\delta \to 0$, $n_1 \to +\infty$) it converges in distribution to the value of the game.

*Remark* 5.5. It can be argued that the expectations of the a-posteriori outcome $\bar{v}$ and of the sampled value of the game $\bar{V}$ have the same monotonicity and convergence properties of the corresponding quantiles $\bar{v}_\alpha(n_1, k_1)$ and $\bar{V}_\alpha(n_1)$.

## 6. Simulations

We now present results of numerical implementation of the randomized technique applied to the output-feedback hide-and-seek game, described in Section 5.

First, we chose a fixed geometry of $m = 10$ candidate treasure points and $s = 2$ sensor points, drawn uniformly from a 50-by-50 square region, and we ran Monte Carlo simulations of the Algorithm 2, with the a-posteriori guarantees described in Theorem 3.5, for increasing values of $n_1$ up to the corresponding a-priori bound (see Theorem 3.4). We set $\bar{n}_2 = 10$, $\beta = 2 \cdot 10^{-5}$, and we repeated the simulations for different values of $\delta$ (or, equivalently, of $k_1$). Figure 4 shows the behavior of the curves $\bar{v}_\alpha(n_1, k_1)$ (for different values of $k_1$) and $\bar{V}_\alpha(n_1)$, as defined in Section 5.1, with $\alpha = 0.9$. For simplicity, we only show curves obtained using the a-posteriori bound in Equation (3.3).

The plots endorse the monotonicity results obtained in Propositions 5.2 and 5.3; furthermore, the curves $\bar{v}_\alpha(n_1, k_1)$ look reasonably "flat", implying that with the choice of $n_1$ that is a few orders of magnitude lower than the a-priori bound, one can obtain a security strategy that has a relatively small value of $\epsilon$ with high probability (see [BBHPD10] for further considerations).

To provide a comparison between the randomized method and the heuristic, we considered a different simulation setup and ran Monte Carlo simulations over 30 different geometries of $m = 10$ candidate treasure points, uniformly randomly distributed in a 50-by-50 square region, and $s = 2$ sensor points placed according to the rectangular heuristic from [SD96]. We applied Algorithm 1, with $\bar{n}_2 = 10$, $\beta = 2 \cdot 10^{-5}$, $\delta = 0.02$, and compared the sampled security value (in terms of quantile $\bar{V}_\alpha(n_1)$) to the cost of the Algorithm 3 (heuristic), as shown in Figure 5.

The heuristic cost represents the seeker's security level, namely the outcome when the seeker uses the heuristic and the hider plays the best response to it, by placing the treasure in the last point to be visited according to the heuristic. Although this value is very good for the hider (i.e., very negative, which corresponds to a large time to find the treasure), we observe a significant gap between this value (the black dotted line) and the SSP curve (the magenta dash-dotted line), especially for large $n_1$. This means that, while the hider can expect a large time until the treasure is found when playing against the heuristic policy, it should not expect such a favorable outcome when playing against the SSP algorithm. Moreover, this particular hiding policy is very fragile because if the seeker learns it, then she can find the treasure in one step. Clearly, the heuristic policy is not a Nash equilibrium. From the seeker's point of view, the heuristic provides a reasonable security level, better than the ETSP cost and with much less computation, because the first part of Algorithm 3 (the 'K steps') allows the exclusion of many points before computing the ETSP path (see Section 4). In general, an advantage of the heuristic with respect to the randomization method is that it does not require the knowledge of the entire geometry to determine the solution, but just the sensor locations and the geometry only at the last step.

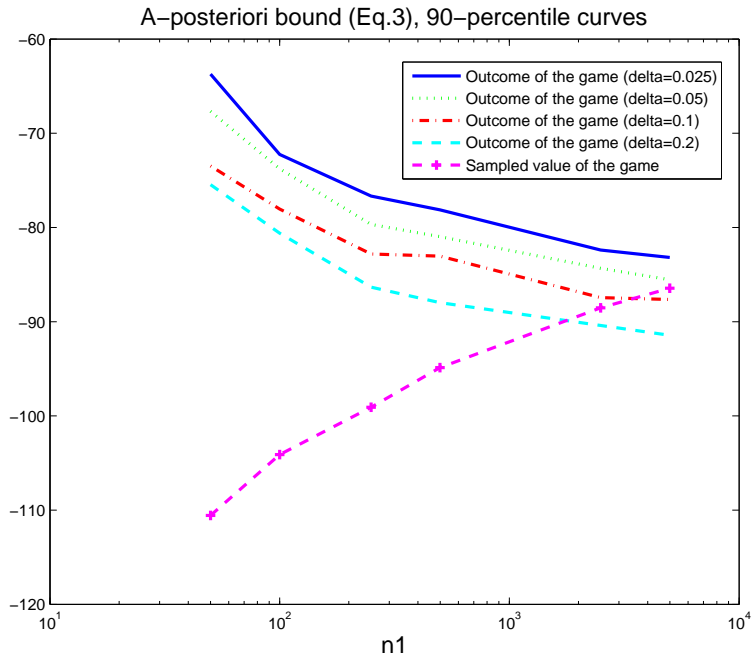

FIGURE 4. Results of Monte Carlo simulations of Algorithm 2 (A-posteriori procedure) over a fixed geometry of points in a planar region, for different confidence levels

It is important to note that the heuristic procedure was constructed assuming that the candidate treasure points are uniformly distributed, but other than that, it has not been optimized to any particular geometry and therefore it should be fairly robust to changes in the positions of the points. In contrast, the SSP technique constructed feedback policies that have been optimized for a specific geometry of the points and can therefore be fragile with respect to changes in the positions of the points away from the positions for which the game was sampled.

## 7. Conclusions and Future Directions

In this paper, we proposed two different techniques to solve the problem of finding a static object in a planar region, when directional measurements are available to the seeker. First, we presented a searching heuristic and characterized bounds on the expected distance covered by the searcher; then we addressed the problem as a large-dimensional game and applied recent results on randomized sampling approach to get security strategies guaranteed with high probability. Simulation results show that, at the cost of performing more computation, the randomized procedure provides a lower (better) security level (with probabilistic guarantees) to the hider, against the one provided by the search heuristic. On the other hand, the heuristic shows its benefits in that it can be implemented very efficiently and its performance does not rely on a specific geometry of the points.

Future work will focus on extending the problem to the continuous case, by assuming that measurements are available in any point of the plane and the object can be put anywhere in the plane; furthermore, the binary information sensor models considered in this paper can be replaced by more sophisticated sensors, providing continuous and noisy measurements.

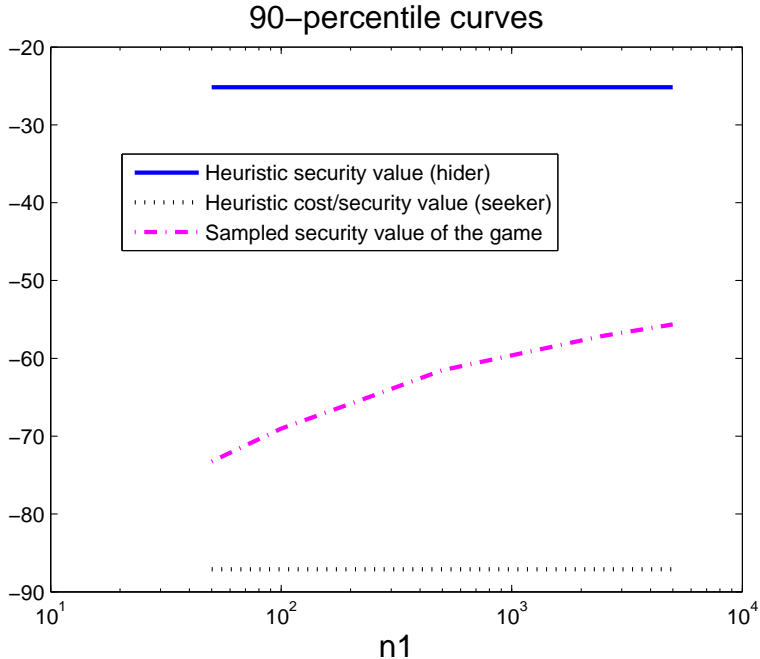

FIGURE 5. Comparison between randomized technique and heuristic, over different geometries. The randomization provides a lower security level than the heuristic, at the cost of performing more computation

## REFERENCES

[AKLV04] Adjler, T., Kozintsev, I., Lienhart, R., and Vetterli, M. (2004). Acoustic source localization in distributed sensor networks. In *Thirty-Eighth Asilomar Conference on Signals, Systems and Computers*, volume 2, 1328–1332. Lausanne, Switzerland.

[BP02] Baum, M.L. and Passino, K.M. (2002). A search-theoretic approach to cooperative control for uninhabited air vehicles. In *AIAA Conference on Guidance, Navigation and Control*. Monterey, CA, USA.

[BSL08] Beck, A., Stoica, P., and Li, J. (2008). Exact and approximate solutions of source localization problems. *IEEE Transactions on Signal Processing*, 56(5), 1770–1778.

[Bel62] Bellman, R. (1962). Dynamic programming treatment of the traveling salesman problem. *J. Assoc. Comput. Mach.*, 9, 61–63.

[BBHPD10] Bopardikar, S.D., Borri, A., Hespanha, J.P., Prandini, M., and Di Benedetto, M.D. (2010). Randomized sampling for large zero-sum games. In *IEEE Conference on Decision and Control (CDC), 2010*, 7675–7680. Atlanta, GA, USA.

[BSBH10] Bopardikar, S.D., Smith, S.L., Bullo, F., and Hespanha, J.P. (2010). Dynamic vehicle routing for translating demands: Stability analysis and receding-horizon policies. *IEEE Transactions on Automatic Control*, 55(11).

[BH11] Bopardikar, S.D., and Hespanha, J.P., (2011). Randomized Solutions to Partial Information Dynamic Games. In *American Control Conference (ACC), June 2011*, San Francisco, CA, USA. Invited paper. To appear.

[BH02] Brass, P. and Heinrich-Litan, L. (2002). Computing the center of area of a convex polygon. Technical report, Freie Universität Berlin, Fachbereich Mathematik und Informatik.

[Bre92] Breiman, L. (1992). *Probability*, volume 7 of *Classics in Applied Mathematics*. SIAM.

[KSIH10] Klein, D.J., Schweikl, J., Isaacs, J.T., and Hespanha, J.P. (2010). On UAV routing protocols for sparse sensor data exfiltration. In *American Control Conference (ACC), 2010*, 6494–6500. Baltimore, MD, USA.

[Res98] Resnick, S. (1998). *A probability path*. Birkhäuser, 1st edition.

[SMH03] Sarmiento, A., Murrieta, R., and Hutchinson, S.A. (2003). An efficient strategy for rapidly finding an object in a polygonal world. In *IEEE/RSJ International Conference on Intelligent Robots and Optimal Systems (IROS)*, 1153–1158.

[Sto07] Stone, L.D. (2007). *Theory of Optimal Search*. Topics in Operations Research Series. INFORMS, MD, USA.

[SD96] Suzuki, A., and Drezner, Z. (1996). The p-center location problem in an area. *Location Science*, Vol. 4, No. 1/2 69–82.

[VMRL03] Valin, J.M., Michaud, F., Rouat, J., and Létourneau, D. (2003). Robust sound source localization using a microphone array on a mobile robot. In *IEEE/RSJ International Conference on Intelligent Robots and Systems*, 1228–1233.

[Was02] Washburn, A.R. (2002). *Search and Detection*. Topics in Operations Research Series. INFORMS, MD, USA.

[1] Department of Electrical and Information Engineering, Center of Excellence DEWS, University of L'Aquila, 67100 L'Aquila, Italy

*E-mail address*:  {alessandro.borri,mariadomenica.dibenedetto}@univaq.it

[2] Center for Control, Dynamical Systems and Computation, University of California at Santa Barbara, CA 93106, USA

*E-mail address*:  {sdbopardikar,hespanha}@ece.ucsb.edu